\documentstyle[epsf,epsfig,12pt]{article}
\begin{document}


\newcommand{\ee}{e$^+$e$^-$}
\newcommand{\ff}{f$_{2}$(1525)}
\newcommand{\bb}{$b \overline{b}$}
\newcommand{\cc}{$c \overline{c}$}
\newcommand{\sbs}{$s \overline{s}$}
\newcommand{\uu}{$u \overline{u}$}
\newcommand{\dd}{$d \overline{d}$}
\newcommand{\qq}{$q \overline{q}$}
\newcommand{\suo}{\rm{\mbox{$\epsilon_{b}$}}}
\newcommand{\loro}{\rm{\mbox{$\epsilon_{c}$}}}
\newcommand{\kos}{\ifmmode \mathrm{K^{0}_{S}} \else K$^{0}_{\mathrm S} $ \fi}
\newcommand{\kol}{\ifmmode \mathrm{K^{0}_{L}} \else K$^{0}_{\mathrm L} $ \fi}
\newcommand{\ko}{\ifmmode {\mathrm K^{0}} \else K$^{0} $ \fi}

\def\tpc{three-particle correlation}
\def\twopc{two-particle correlation}
\def\ksks{K$^0_S$K$^0_S$}
\def\ee{e$^+$e$^-$}
\def\ff{f$_{2}$(1525)}

\begin{center}
{\LARGE Hadronic Decays of $J/\psi$(1$^{3}S_{1}$) and  
$\psi^{\prime}$(2$^{3}S_{1}$) Through Vitrual Photons~}
\end{center}

\begin{center}

{Kamal~K.~Seth}

{Northwestern University, Evanston, Illinois 60208}
\end{center}



\date{January 11, 2004}

\begin{abstract}

The latest data for $J/\psi$ and $\psi^{\prime}$ leptonic decays,
and the latest measurement of the R--parameter are used to show that
the present summary values of the branching ratios of the virtual photon
mediated hadronic
decays of both  $J/\psi$ and $\psi^{\prime}$ are overestimated.
The current best results are 
$B(J/\psi \rightarrow \gamma^{*} \rightarrow h) = (13.3 \pm 0.3)$\%,
and  
$B(\psi^{\prime} \rightarrow \gamma^{*} \rightarrow h) = (1.65 \pm 0.10)$\%.

\end{abstract}



The  $J/\psi$(3097) and $\psi^{\prime}$(3686) 1$^{--}$ states of charmonium
couple to leptons only through virtual photons, and the branching
ratios $B(J/\psi,\psi^{\prime} \rightarrow \gamma^{*} \rightarrow hadrons)$
are among the largest decay ratios for both. These ratios were first 
reported by Boyarski {\it et al.}~\cite{boyarski} and Luth 
{\it et al.}~\cite{luth} from the 1975 measurements with the Mark I 
detector at SLAC. The Review of Particle Properties (PDG) has continued
to quote these results ever since~\cite{pdg2002}.  The quoted values are:
\begin{equation}
B(J/\psi \rightarrow \gamma^{*} \rightarrow h) = (17.0 \pm 2.0)\% \, ,
\end{equation}
\begin{equation}
B(\psi^{\prime} \rightarrow \gamma^{*} \rightarrow h) = (2.9 \pm 0.4)\% \, .
\end{equation}
Both these results were obtained by using the relation
\begin{displaymath}
B(J/\psi,\psi^{\prime} \rightarrow \gamma^{*} \rightarrow h) =
{\sigma_{h}(nonresonant) \over \sigma_{l}(nonresonant)} 
\times B_{l}(J/\psi,\psi^{\prime} \rightarrow l^{+}l^{-}) \,
\end{displaymath}
\begin{equation}
~~~~~~~~~~~~~~~~~~~~~~~~~~~\equiv R(nonresonant) \times B(J/\psi,\psi^{\prime} \rightarrow l^{+}l^{-}) \, .
\end{equation}

Boyarski {\it et al.}~\cite{boyarski} reported 
$B(J/\psi \rightarrow \mu^{+}\mu^{-})=(6.9 \pm 0.9)$\%.
Using the value $R = 2.5 \pm$ 0.3 measured by 
Augustin {\it et al.}~\cite{augustin}, they obtained 
$\Gamma(J/\psi \rightarrow \gamma^{*} \rightarrow h) = (12 \pm 2)$ keV,
or equivalently the PDG result of Eq. (1).

\begin{figure}[hbt]
\vspace*{-0.3cm}
\label{fg:fits}
\begin{center}
\includegraphics[width=6.in]{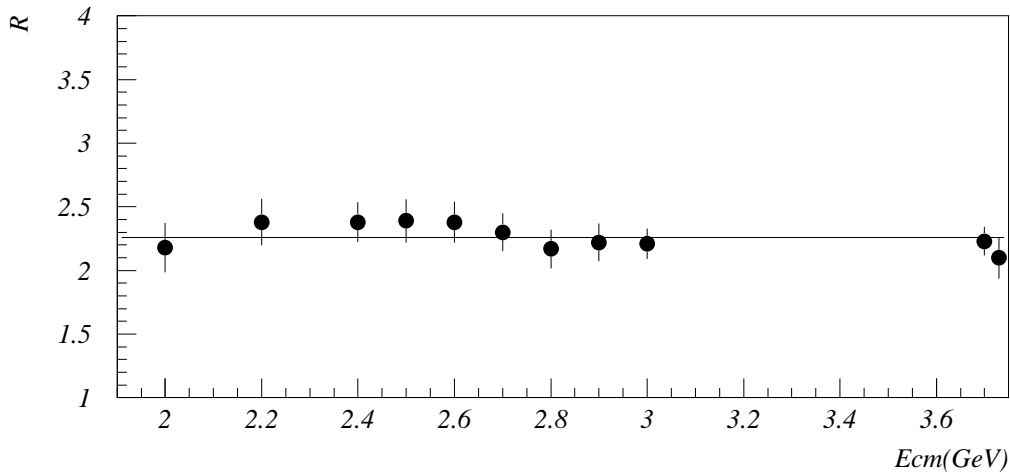}
\end{center}
\vspace*{-0.6cm}
\caption{$R$ measurements in the c.m. energy range from 2.0 to 3.73 GeV
from BES~\cite{bes} with best fit, $R$=2.26$\pm$0.04.}
\end{figure}

More recently, much more precise result for both $R(nonresonant)$
and the leptonic branching ratio have become available. 
BES~\cite{bes} has reported high precision results for $R$ in the
region 2--5 GeV, and PDG~\cite{pdg2002} itself has fitted
and summarized the latest results for the leptonic branching
ratios. The $R$ values measured by BES~\cite{bes} are constant
in the region 2.0 to 3.73 GeV (Fig.~1). The result of our fit to these
data is
\begin{equation}
R = 2.26 \pm 0.04 \, 
\end{equation}
which is very close to the theoretical expectations for three flavors,
$R$=2.13 (correct to third order)~\cite{pdg2002}. 
The average of $B(J/\psi \to e^{+}e^{-})$ and $B(J/\psi \to \mu^{+}\mu^{-})$
is~\cite{pdg2002}
\begin{equation}
B(J/\psi \to l^{+}l^{-}) = (5.90 \pm 0.09)\% \, .
\end{equation}
These leads to 
\begin{equation}
B(J/\psi \rightarrow \gamma^{*} \rightarrow h) = (13.3 \pm 0.3)\% \, ,
\end{equation}
a result which represents a large improvement over Eq. (1).

For $\psi^{\prime}$(2$^{3}S_{1}$) Luth {\it et al.}~\cite{luth} reported
$B(\psi^{\prime} \to e^{+}e^{-}) = (0.93 \pm 0.16)$\% and apparently
used $R$=3.1 to obtain 
$B(\psi^{\prime} \rightarrow \gamma^{*} \rightarrow h) = (2.9 \pm 0.4)$\%.
The PDG listing is 
$B(\psi^{\prime} \to l^{+}l^{-}) = (0.73 \pm 0.04)$\%~\cite{pdg2002}.
Using fitted value of $R$ in Eq. (4), we obtain
\begin{equation}
B(\psi^{\prime} \rightarrow \gamma^{*} \rightarrow h) = (1.65 \pm 0.10)\% \, ,
\end{equation}
a result which is significantly different from that of Eq. (2).

Since hadronic decays through virtual photons are included in the total
hadronic decays, these revised values do not change the branching
ratios, 
$B(J/\psi,\psi^{\prime} \rightarrow h)$, but they do alter the 
estimates of $B(J/\psi,\psi^{\prime} \rightarrow ggg)$.

This work was supported by the U.S. Department of Energy.

\end{document}